\pdfoutput=1
\documentclass[10pt,logo,copyright]{nvidiatechreport}
\usepackage[numbers,sort]{natbib}
\definecolor{citecolor}{HTML}{0071BC}
\definecolor{linkcolor}{HTML}{ED1C24}

\makeatletter
\@ifpackageloaded{hyperref}{}{
    \usepackage[pagebackref=false,breaklinks=true,colorlinks,citecolor=citecolor,linkcolor=linkcolor,bookmarks=false]{hyperref}
}
\makeatother

\usepackage{epsfig}
\usepackage{graphicx}
\usepackage{amsmath,mathtools}
\usepackage{amssymb}
\usepackage{amsthm}
\usepackage{nicefrac}       %
\usepackage{microtype}      %
\usepackage{amsmath} 
\usepackage{float}
\usepackage{amsfonts}
\usepackage{bm}
\usepackage{enumitem}
\usepackage{mathtools}
\usepackage{multirow}
\usepackage{siunitx}

\usepackage{fontawesome}
\usepackage{pifont}
\usepackage{color}
\usepackage[algo2e,inoutnumbered,linesnumbered,algoruled,vlined]{algorithm2e}
\usepackage{algpseudocode}
\usepackage{tabularx}
\usepackage{url}
\usepackage{bbm}
\usepackage{threeparttable}
\usepackage{tikz}
\usepackage{wrapfig}
\usetikzlibrary{backgrounds}
\usetikzlibrary{decorations.pathreplacing}
\usepackage[outline]{contour}
\usepackage{subcaption}
\usepackage{wrapfig}

\newcommand{\parahead}[1]{\vspace{1.5mm}\noindent\textbf{#1.}\ }

\theoremstyle{definition}

\newcommand{\RR}{\mathbb{R}}

\newcommand{\SpS}{\mathbb{S}}

\newcommand{\trans}{\mathbf{T}}

\newcommand{\loss}{\mathcal{L}}

\DeclarePairedDelimiterX{\infdivx}[2]{(}{)}{%
  #1\;\delimsize\|\;#2%
}

\definecolor{darkblue}{RGB}{49,130,189}
\definecolor{stanfordgrey}{RGB}{46,45,41}
\definecolor{cardinalred}{RGB}{253,141,60}

\newcommand{\Image}{{\mathbf{I}}}

\newcommand{\Feature}{{\mathbf{F}}}
\newcommand{\Geometry}{{\mathcal{G}}}

\newcommand{\gmu}{{\bm{\mu}}}

\newcommand{\gscale}{{\bm{s}}}
\newcommand{\gquat}{{\mathbf{q}}}

\newcommand{\Attr}{\mathbf{A}}

\newcommand{\encw}{{\bm{\phi}}}     %

\makeatletter
\DeclareRobustCommand\onedot{\futurelet\@let@token\@onedot}
\def\@onedot{\ifx\@let@token.\else.\null\fi\xspace}

\makeatother

\usepackage[capitalize]{cleveref}

\crefname{section}{\S}{\S\S}
\crefname{subsection}{\S}{\S\S}
\crefname{conj}{Conj.}{Conj.}

\Crefname{assumption}{\textbf{H}\hspace{-3pt}}{\textbf{H}\hspace{-3pt}}
\crefname{assumption}{\textbf{H}}{\textbf{H}}

\crefname{algorithm}{Alg.}{Alg.}
\crefname{assumption}{\textbf{H}}{\textbf{H}}
\crefname{equation}{Eq.}{Eq.}
\crefname{definition}{Dfn.}{Dfn.}
\crefname{lemma}{Lemma}{Lemma}
\crefname{dfn}{Dfn.}{Dfn.}
\crefname{thm}{Thm.}{Thm.}
\crefname{tab}{Tab.}{Tab.}
\crefname{fig}{Fig.}{Fig.}
\crefname{table}{Tab.}{Tab.}
\crefname{figure}{Fig.}{Fig.}

\definecolor{mygreen}{RGB}{159, 200, 59}
\definecolor{myred}{RGB}{223, 135, 102}

\newcommand{\MethodName}{\textit{Asset Harvester}\xspace}
\newcommand{\PipeName}{\textit{Asset Harvester}\xspace}

\definecolor{policy_green}{rgb}{0.2, 0.59, 0.2}
\definecolor{policy_yellow}{rgb}{0.8, 0.8.0, 0.2}

\usepackage{amsmath,amsfonts,bm,mathtools}

\def\eqref#1{equation~\ref{#1}}

\def\1{\bm{1}}

\def\rmK{{\mathbf{K}}}

\def\rmQ{{\mathbf{Q}}}

\def\rmV{{\mathbf{V}}}

\DeclareMathAlphabet{\mathsfit}{\encodingdefault}{\sfdefault}{m}{sl}
\SetMathAlphabet{\mathsfit}{bold}{\encodingdefault}{\sfdefault}{bx}{n}

\def\@onedot{\ifx\@let@token.\else.\null\fi\xspace}

\renewcommand{\MethodName}{\textit{Instant NuRec}\xspace}
\renewcommand{\PipeName}{\textit{Instant NuRec}\xspace}

\title{Instant NuRec: Feed-Forward 3D Gaussian Reconstruction for Driving Scene Simulation}
\author{NVIDIA\textsuperscript{\textdagger} \\
\url{https://research.nvidia.com/labs/sil/projects/instant-nurec}}

\begin{document}
\maketitle
\begingroup
\renewcommand{\thefootnote}{\textdagger}
\footnotetext{The full list of contributors is provided in Appendix~\ref{sec:contributors}.}
\endgroup

\noindent
\begin{minipage}{\linewidth}
    \centering
    \includegraphics[width=\linewidth]{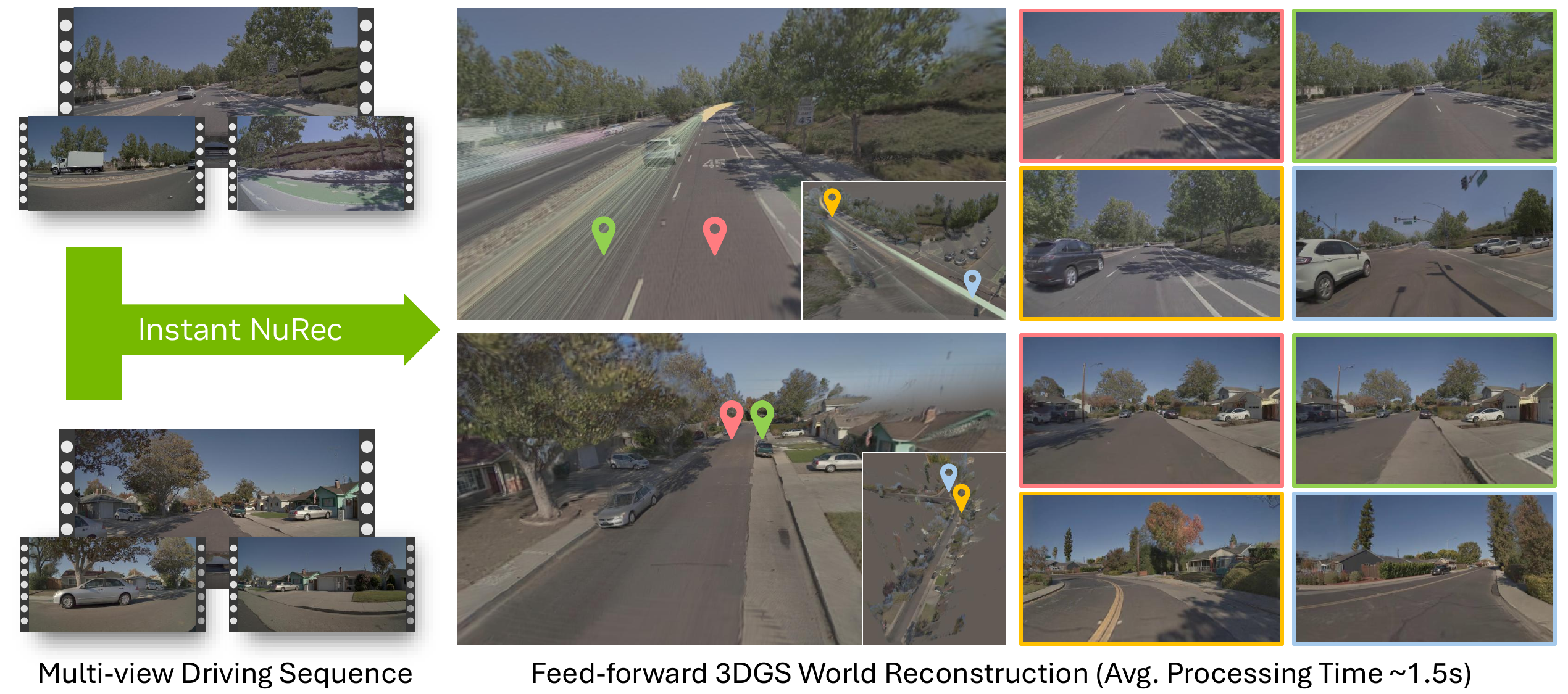}
    \captionof{figure}{\textbf{\MethodName} ingests a multi-camera driving sequence and, in a single feed-forward pass, produces a layered 3DGS world with static, dynamic, and sky components that the user can freely navigate spatially and temporally. For each scene (top and bottom), the location markers indicate the vehicle positions used to produce the renderings in the right columns.}
    \label{fig:teaser}
\end{minipage}
\vskip12pt

\begin{abstract}
3D simulation platforms are critical for autonomous driving because they enable end-to-end policy evaluation, thereby reducing development costs and improving safety.
In recent years, neural simulation has become predominant, with methods such as NuRec~\cite{nurec_website} playing a central role; however, these methods remain relatively slow and typically require per-scene tuning.
In this work, we present \PipeName{}, a feed-forward neural reconstruction model that turns a short multi-view driving log into a fully simulatable 3D Gaussian Splatting (3DGS) world in a single forward pass.
The model accepts multi-view input from a calibrated camera rig and emits a layered output consisting of static and dynamic 3DGS layers, a sky cubemap, and per-camera ISP corrections, while providing native support for non-pinhole camera models via 3DGUT~\cite{wu20253dgut}.
It reconstructs a 10--20-second multi-camera scene in roughly 1.5 seconds and achieves a PSNR on the Waymo Open Dataset that is 2.01\,dB above the strongest evaluated baseline.
\PipeName{} is deeply integrated into NuRec and is compatible with AlpaSim~\cite{alpasim_2025} for closed-loop simulation.
\end{abstract}

{\hbadness=10000\abscontent\par}

\section{Introduction}
\label{sec:introduction}

Closed-loop simulation has become a central tool for developing and validating autonomous driving (AD) stacks: it allows planners and perception modules to be tested against rare, safety-critical, or otherwise hard-to-collect interactions in a controllable environment.
A faithful simulator must combine three ingredients: \textit{(i)} photorealistic rendering of the surrounding world, \textit{(ii)} an explicit, physics-simulation-ready decomposition of the scene into a static background and individually editable dynamic agents, and \textit{(iii)} the ability to freely re-pose the ego camera and re-time the scene so that counterfactual rollouts can be evaluated.
Recent advances in 3D Gaussian Splatting (3DGS)~\citep{kerbl2023gaussian} have made high-quality rendering computationally inexpensive, while per-scene optimization frameworks such as OmniRe~\citep{chen2025omnire} have shown that the resulting reconstructions can faithfully reproduce a real driving log.
However, per-scene optimization is fundamentally a bottleneck to scaling: each new clip requires hours of computation, careful hyperparameter tuning, and a large set of auxiliary inputs (LiDAR sweeps, fitted poses, and semantic masks).
This cost makes it impractical to ingest the millions of clips that modern AD fleets collect every day.

In this paper, we introduce \MethodName, a feed-forward neural reconstruction model that turns a short multi-view driving log into a fully simulatable 3DGS world in a single forward pass.
The input is a sequence of images captured by a rig with 1--5 calibrated cameras.
A single forward pass through \MethodName produces \textit{(i)} a static 3DGS layer that captures roads, buildings, and other rigid background elements; \textit{(ii)} a dynamic 3DGS layer with piecewise-linear trajectories that represents moving vehicles and vulnerable road users; and \textit{(iii)} a cubemap that models the distant sky.
The output can be exported in USDZ format and consumed directly by closed-loop simulation frameworks such as AlpaSim~\citep{alpasim_2025}, allowing a change in clip selection to be reflected in the simulator within seconds rather than hours.

\MethodName{} couples a single shared encoder with a set of lightweight task-specific decoders.
The encoder is an alternating-attention Vision Transformer~\citep{dosovitskiy2020vit} that interleaves intra-image and cross-image attention to fuse information across the input cameras and over time.
Its shared features drive several decoders that predict dense depth, normals, semantics, a sky cubemap, and per-camera color corrections. The predicted depth lifts a set of query points into world space to anchor the Gaussians, the semantic head separates the static background from dynamic agents, and a Gaussian head supplies the remaining shape and appearance attributes needed to complete the layered output.
Crucially, a motion head directly recovers piecewise-linear actor trajectories for the query points, so the dynamic layer requires no per-actor cuboid tracks at reconstruction time.
We train the model in three stages that deliberately decouple geometry supervision from rendering supervision, keeping large-scale training on $\sim$40K internal driving clips efficient.

In short, \MethodName{} turns a multi-camera driving log into a complete 3DGS world with static, dynamic, and sky components in a single feed-forward pass, ready for direct use in closed-loop simulation.
Across novel-view synthesis, geometry, and downstream simulator integration, it closes a large fraction of the quality gap to per-scene optimization while requiring $10^3$--$10^4\!\times$ less computation per clip, making large-scale reconstruction of driving logs practical.
Our code is available at \url{https://github.com/nvidia/instant-nurec}, with additional documentation at \url{https://docs.nvidia.com/nurec/index.html}.

\section{Related Work}
\label{sec:related_work}

\parahead{Per-scene neural reconstruction for driving}
Neural Radiance Fields (NeRFs)~\citep{mildenhall2020nerf} and 3DGS~\citep{kerbl2023gaussian} have set the standard for photorealistic novel-view synthesis, and a substantial body of follow-up work has specialized them to outdoor driving scenes.
Early urban radiance fields scaled reconstruction to street and city scales using posed RGB and LiDAR~\citep{rematas2022urbanradiance,tancik2022blocknerf}, while Neural Scene Graphs~\citep{ost2021neural} introduced the now-standard decomposition of a driving scene into a static background and a set of rigidly moving foreground actors.
Later methods pushed this toward self-supervised dynamics and sensor-realistic simulation~\citep{yang2023emernerf,tonderski2024neurad} and replaced the volumetric field with explicit Gaussians for faster, editable reconstruction~\citep{zhou2024drivinggaussian,yan2024streetgaussians,chen2023pvg}.
OmniRe~\citep{chen2025omnire} extended this decomposition to non-rigid pedestrians and cyclists, while 3DGRT~\citep{moenneloccoz20243dgrt} and 3DGUT~\citep{wu20253dgut} introduced Gaussian renderers that natively support distorted cameras and secondary rays, removing the rectified-pinhole assumption.
These methods produce extremely high-quality reconstructions but require minutes to hours of per-clip optimization and a rich set of auxiliary inputs (LiDAR, multi-traversal poses, semantic masks, and cuboid tracks).
\MethodName preserves the same output format, namely a layered 3DGS world consumable by NuRec~\citep{nurec_website,carla_nurec_docs} and AlpaSim~\citep{alpasim_2025}, but amortizes the optimization cost into a single feed-forward pass.

\parahead{Feed-forward 3D reconstruction}
Feed-forward 3D models have rapidly progressed from single-object generation~\citep{liu2023zero1to3,shi2023mvdream,wang2023imagedream,tang2024lgm,xiang2024trellis,xiang2025trellis2,zhao2025hunyuan3d2,sam3dteam2025sam3d3dfyimages} to scene-level prediction.
At the scene level, generalizable Gaussian models predict per-pixel Gaussians from sparse posed views in a single pass using probabilistic depth~\citep{charatan2024pixelsplat}, plane-sweep cost volumes~\citep{chen2024mvsplat}, large transformer decoders~\citep{zhang2024gslrm}, or monocular depth priors~\citep{xu2024depthsplat}. In parallel, another line of work regresses pointmaps or Gaussians in a pose-free manner directly from uncalibrated images~\citep{wang2024dust3r,ye2024noposplat,jiang2025anysplat}.
This paradigm has recently been specialized to dynamic driving scenes: STORM~\citep{yang2025storm} and DrivingForward~\citep{tian2024drivingforward} predict static and dynamic Gaussians from surround-view video, and a wave of contemporaneous models extends this to pose-free, tracker-free, and off-road settings~\citep{chen2025dggt,yu2026recondrive,yu2026streetforward,wang2026ground4d}.
A separate line of feed-forward 3DGS work decouples Gaussian prediction from input pixels: instead of emitting one Gaussian per pixel, TokenGS~\citep{tokengs2026} regresses Gaussians from a fixed set of learnable tokens, and related query- or scene-token formulations~\citep{peng2026uniquer,itkin2026globalsplat} similarly unbind the primitive count from image resolution and view count.
Unlike these reconstruction-focused models, which target bare RGB Gaussians, \MethodName{} emits a complete, simulation-ready layered 3D world.

\parahead{Driving world models}
A complementary and rapidly growing line of work learns generative models that synthesize plausible driving observations conditioned on the past and on control signals such as ego trajectory, layout, or text.
Early latent-diffusion models~\citep{hu2023gaia1,wang2023drivedreamer,gao2023magicdrive,wen2023panacea} generate controllable single- or multi-view street video. Subsequent work improves fidelity and controllability, extends the prediction horizon~\citep{wang2023drivingfuture,gao2024vista,zhao2024drivedreamer2,russell2025gaia2,hassan2024gem}, adopts autoregressive video-GPT or diffusion backbones for long-range rollouts~\citep{hu2024drivingworld,zhang2025epona}, and culminates in large world foundation models trained across driving and other embodiments~\citep{agarwal2025cosmos,ren2025cosmosdrivedreams}.
These models are powerful content generators but, by construction, emit 2D pixels rather than an explicit 3D state: they cannot be freely re-posed by a planner and do not directly yield a simulatable scene.
A second line therefore uses generative video priors to \emph{repair or densify} neural reconstructions, synthesizing novel-trajectory frames offline as a data machine~\citep{zhao2024drivedreamer4d,mao2025dreamdrive} or removing ghosting and floater artifacts online as an inference-time enhancer~\citep{ni2024recondreamer,wu2025difix3d,zhang2026diffusionharmonizer}. Intermediate variants are conditioned on point clouds or pseudo-renders~\citep{yan2024streetcrafter,wang2024freevs}.
Most recently, joint formulations such as the Xiaomi EV world model~\citep{zhou2026xiaomiwm} tightly couple a feed-forward reconstructor with a video generator, allowing geometry to anchor generation while the generator fills unobserved regions.
\MethodName{} occupies the reconstruction side of this framework: it does not invent content but emits an explicit, navigable 3D world in a single pass, making it complementary to generative world models and a natural, fast 3D backbone for the generative-repair and joint-model lines.

\section{Method}
\label{sec:method}

\begin{figure}[htbp]
    \centering
    \includegraphics[width=\linewidth]{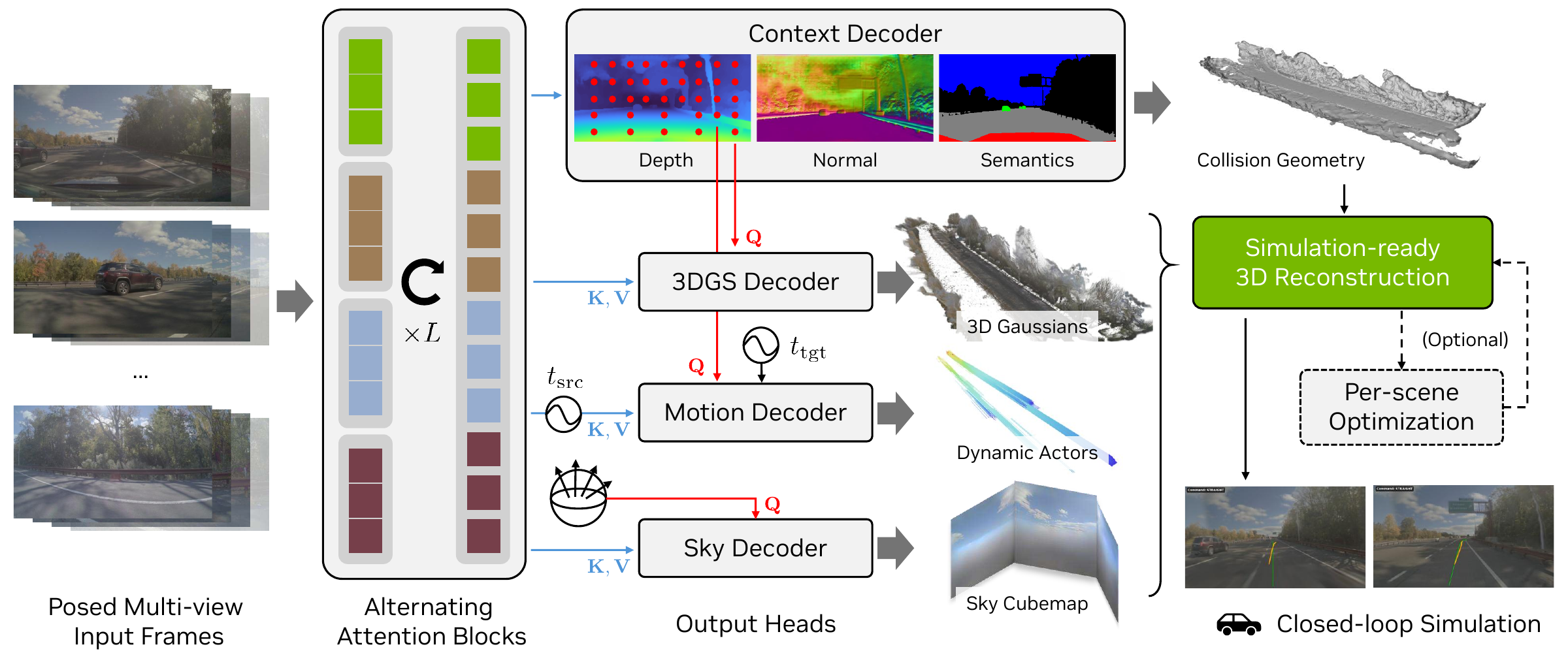}
    \caption{\textbf{Pipeline overview.} Multi-view driving images are tokenized into patches and processed by an alternating-attention ViT encoder. Several decoder heads share the resulting latent features and produce depth maps, semantic labels, motion estimates, a sky cubemap, and 3DGS attributes. Optionally, the output can be further optimized on a per-scene basis and used for downstream simulation tasks.}
    \label{fig:pipeline}
\end{figure}

At a high level, \MethodName takes a short multi-view driving log as input and produces a layered 3DGS world in a single feed-forward pass.
We first describe the problem formulation (input/output representation) in \Cref{sec:io}, then the shared encoder in \Cref{sec:encoder}, and finally the decoder heads that produce the renderable representation in \Cref{sec:decoders}.
A schematic of the pipeline is shown in \Cref{fig:pipeline}.

\subsection{Problem Formulation}
\label{sec:io}

\parahead{Input}
The model accepts $V\!\times\!T$ RGB images, where $V$ is the number of available cameras and $T$ is the number of temporal frames sampled from the source camera videos at 2--4\,Hz.
Each image $\Image_{v,t}$ is paired with its 6-DoF pose $\trans_{v,t}\!\in\!SE(3)$ and a vector of camera-intrinsic parameters $\bm{\kappa}_{v,t}$.
Optional cuboid tracks for dynamic actors can be supplied at inference time to further calibrate the trajectories of the dynamic Gaussians.

\parahead{Output}
\MethodName{} emits a layered representation:
\begin{itemize}[leftmargin=1.2em]
    \item \textbf{Static layer} $\Geometry^s$, a set of $N_s$ static Gaussians, each parameterized by attributes $(\gmu_i,\gscale_i,\gquat_i,\alpha_i,c_i,\bm{n}_i,\ell_i)$, with world-space position $\gmu\!\in\!\RR^3$, scale $\gscale\!\in\!\RR^3$, rotation quaternion $\gquat$, opacity $\alpha\!\in\![0,1]$, color $c\!\in\!\RR^3$, world-space normal $\bm{n}\!\in\!\SpS^2$, and a semantic class $\ell$ (road/non-road) used by downstream simulators. The quaternion primarily parameterizes the Gaussian's rotation for view rendering, whereas the normal is used primarily to extract surfaces for simulation.
    \item \textbf{Dynamic layer} $\Geometry^d$, a set of $N_d$ dynamic Gaussians with the same per-Gaussian attributes as the static layer, except that the position $\gmu_i$ is replaced by a piecewise-linear trajectory $\gmu(t)$ defined by knots $\{(t_k,\gmu_k)\}_{k=1}^{3}$. For $t \in [t_k,t_{k+1}]$, the trajectory is evaluated as $\gmu(t)\!=\!(1\!-\!\lambda)\,\gmu_k\!+\!\lambda\,\gmu_{k+1}$ for $\lambda\!=\!(t\!-\!t_k)/(t_{k+1}\!-\!t_k)$. The opacity is gated by a smooth fade $f(x)\!=\!\exp(-x^{10})$ around $t_1$ and $t_3$.
    \item \textbf{Sky cubemap} $\Image^{\mathrm{sky}}\!\in\!\RR^{6\times H\times W\times 3}$, which represents the distant background.
    \item \textbf{ISP affine transforms} $\{\Attr_v\}\!\in\!\RR^{V\times 3\times 4}$, applied to rendered colors before the photometric loss to absorb residual per-camera color/exposure mismatch.
\end{itemize}

\subsection{Shared Alternating-Attention Encoder}
\label{sec:encoder}

We largely follow the backbone design of Depth-Anything-3~\cite{lin2025depth}, which repurposes the DINOv2~\cite{oquab2023dinov2} architecture for multi-view inputs.
Compared with other geometry backbones, it produces sharp and accurate depth maps without requiring an additional image tokenizer.

In our model, the encoder $\encw$ jointly tokenizes all input images and produces a shared latent $\Feature\!\in\!\RR^{N\times C}$, where $N$ is the total number of tokens and $C$ is the feature dimension.
Each image is split into non-overlapping $14\!\times\!14$ patches, each of which is linearly projected to a token of dimension $C$.
A per-image \emph{class token}, derived from the camera's 6-DoF pose and field of view using sinusoidal positional encoding, is prepended to provide a global summary of the view and inject pose information directly into attention.
The combined sequence is processed by a stack of $L$ alternating attention blocks that interleave \textit{intra-image} self-attention and \textit{cross-image} self-attention.
This interleaving design enables effective information exchange between patches in different views while maintaining computational efficiency in layers that apply intra-image attention.

\subsection{Decoding Heads}
\label{sec:decoders}

Several decoding heads operate on the shared latent features to produce the final output.
These decoders are lightweight, so adding an output head incurs little additional cost at inference.

\parahead{Context-view decoders}
We use \emph{context views} to denote the input images observed by the model, as opposed to held-out target views used for rendering supervision and evaluation.
The context-view decoders produce dense, pixel-aligned scene attributes for these input views.
We use a set of DPT~\cite{ranftl2021vision} fusion heads to fuse latent features from multiple layers into attribute maps.
Specifically, we attach three DPT heads: one each for depth, normals, and semantic logits.
The depth branch follows the Depth-Anything-3 model, but we remove the confidence channel because reconstruction tasks generally require dense predictions over the full image.
For the semantic logits, we predict four classes: road, movable (vehicles and pedestrians), sky, and ego-car.

\parahead{3DGS decoder}
Most existing feed-forward reconstruction methods emit one Gaussian at every pixel.
While this usually provides strong reconstruction quality, the number of Gaussians can be large and their pixel-aligned geometry may not reflect the underlying scene structure.
In our design, we decouple the Gaussians from the latent features by selecting a set of query points from the lifted depth map.
Each query point in $\rmQ$ is a 3D position lifted from the predicted depth map and embedded by a lightweight learned projection.
The query points cross-attend to keys $\rmK$ and values $\rmV$ derived from the latent feature $\Feature$, and the decoder outputs the per-query Gaussian attributes, including scale, rotation, and opacity.
After decoding, the Gaussians from all queries are combined and can then be rendered by 3DGUT~\citep{wu20253dgut}.

\parahead{Motion decoder}
The motion decoder takes the same set of query points $\rmQ$ as input and predicts 3D displacements to the target timestamps $t_{\textrm{tgt}}$.
For computational efficiency, we jointly predict two 3D displacements, one for each target timestamp, yielding a 6-D regression target.
The motion decoder combines elements of V-DPM~\cite{sucar2026v} and 4RC~\cite{luo20264rc}: a shallow Transformer operates on the encoder features and is modulated by the source timestamps $t_{\textrm{src}}$. The Transformer's LayerNorms are AdaLN modules~\cite{peebles2023scalable} conditioned on the target timestamps $t_{\textrm{tgt}}$, informing the head which source-to-target motion $t_{\textrm{src}} \rightarrow t_{\textrm{tgt}}$ to predict.
For each source timestamp $t_{\textrm{src}}$, we select the immediately preceding and following timestamps from the same camera as targets.
Given displacement predictions $(\Delta_-, \Delta_+)$, the three final 3D positions of the dynamic layer are $\gmu_1 = \gmu + \Delta_-$, $\gmu_2 = \gmu$, and $\gmu_3 = \gmu + \Delta_+$, where $\gmu$ is the 3D position lifted from the depth map.
Unlike STORM~\cite{yang2025storm}, in which all Gaussians are considered dynamic, we use the same query set throughout and subsequently apply a semantic-head mask, assigning only points classified as \emph{movable} to the dynamic layer and the remainder to the static layer.

\parahead{Query point selection}
Query points anchor the predicted Gaussians in space, so their selection strongly affects reconstruction quality.
We consider two strategies:
(i) \textbf{Dense}, where every pixel in the context images is a query point; and
(ii) \textbf{Selective}, which uses strided sampling and cross-attention to sparsify queries and includes a dedicated \emph{road branch} for road-dominated regions.
In Selective, query points, each associated with full-resolution per-pixel signals, are sampled with stride $s$ and grouped into $p\!\times\!p$ token clusters.
Each query token cross-attends to the latent feature map (with QK normalization and layer scale for stability) and predicts a fixed cluster of $M\!=\!p^2$ Gaussians, whose attributes are gathered from the corresponding group of signals.
For example, with $s\!=\!2$ and $p\!=\!8$, a single $448\!\times\!784$ context view yields $1{,}372$ tokens and roughly $88$K Gaussians with the \textbf{Selective} strategy. This represents a $4\times$ reduction from the $\sim\!351$K produced by \textbf{Dense} while empirically retaining most of the reconstruction quality.

\parahead{Sky decoder}
The sky decoder maps the shared latent $\Feature$ to the distant-background cubemap $\Image^{\mathrm{sky}}$.
For each cubemap pixel, we form a query by sinusoidally encoding its world-space ray direction. The query cross-attends to encoder features augmented with per-camera ray embeddings so that each pixel can preferentially borrow color from the input view that observed a similar direction.
A DPT-style fusion head upsamples the cross-attention output to full cubemap resolution.

\parahead{ISP decoder}
This decoder predicts a per-camera $3\!\times\!4$ affine RGB transform, analogous to appearance compensation in neural rendering~\citep{martinbrualla2021nerfw}, that absorbs residual differences in white balance, gamma, and vignetting between cameras observing the same scene.
A learned token cross-attends to the deepest encoder features, grouped by camera index, and a final linear projection emits the affine matrix and bias.
The transform is initialized to the identity and applied to rendered colors before the photometric loss is computed.
At reconstruction time, downstream simulators may either apply the transform to match the training rig's color response or omit it to render in the canonical color space.

\parahead{Forward-pass summary}
Putting the pieces together, the encoder $\encw$ maps the input clip to a shared latent representation $\Feature$, from which the context DPT heads predict dense depth, normals, and semantics.
Query points $\rmQ$ are lifted from the depth map using either the Dense or Selective strategy. The 3DGS decoder cross-attends $\rmQ$ to $\Feature$ and assembles static Gaussians $\Geometry^s$ with scale, rotation, opacity, color, normal, and semantic labels. The motion decoder uses the same queries $\rmQ$ to predict displacements; only queries subsequently masked as \emph{movable} are assigned to dynamic Gaussians $\Geometry^d$ with piecewise-linear trajectories.
The sky and ISP heads produce $\Image^{\mathrm{sky}}$ and per-camera affine transforms $\{\Attr_v\}$ in parallel; no iterative optimization is performed at inference.

\subsection{Long-clip Processing via Chunk Merging}
\label{sec:chunking}

Because the model is trained on short temporal windows for memory and computational efficiency, reconstructing an entire driving log requires processing it in overlapping chunks and merging the per-chunk predictions into a single global field.

\parahead{Per-chunk pruning}
Within each chunk, we drop Gaussians whose opacity is below a small threshold ($\alpha\!<\!10^{-2}$) and transform the surviving primitives into a global world frame using the chunk-level pose.
In practice, this can reduce the number of Gaussians while maintaining rendering quality.

\parahead{Frustum-ownership pruning}
A Gaussian predicted in one chunk, particularly at a great distance from the chunk's cameras, often lands inside another chunk's frustum, where it competes with that chunk's own closer, better-conditioned Gaussians and creates floaters.
We therefore apply a \emph{frustum-ownership} test: a Gaussian $g$ from chunk $i$ is kept only if no other chunk $j$ observes it from substantially closer range.
Specifically, define $D_{gi}$ as the Euclidean distance from the Gaussian's center to the optical center of its closest camera in chunk $i$, with $D_{gi}\!=\!\infty$ if $g$ lies outside chunk $i$'s frustum pyramid. We retain $g$ from chunk $i$ if and only if
\begin{equation}
D_{gi}\;\le\;\min_{j}\,D_{gj} + \delta
\label{eq:frustum-mask}
\end{equation}
where $\delta$ is a small tolerance that keeps Gaussians on the boundary between chunks.
Intuitively, each Gaussian is ``owned'' by the chunk that observed it at the closest range; downstream chunks see only their own near-field reconstruction and inherit the far field from their closer neighbors.
Because dynamic Gaussians are active only over a short time window and rarely overlap with another chunk's active interval, we leave the dynamic layer untouched and apply the pruning to the static layer only.

\section{Training}
\label{sec:training}

We describe the training objective and curriculum in \Cref{sec:curriculum} and the implementation details in \Cref{sec:impl}.

\subsection{Losses and Training Curriculum}
\label{sec:curriculum}

We train the model with three groups of objectives:
\begin{equation}
\loss = \loss_{\text{context}} + \loss_{\text{motion}} + \loss_{\text{render}},
\end{equation}
where
\begin{itemize}[leftmargin=1.2em]
    \item $\loss_{\text{context}}$ directly supervises the outputs of the context decoders. Attributes such as depth, normals, and semantic logits are pixel-aligned and have corresponding ground truth. We supervise these attributes with $L_1$, cosine similarity, and cross-entropy losses, respectively.
    \item $\loss_{\text{motion}}$ includes forward and backward 3D-flow losses, supervised by 3D scene flow derived from cuboid tracks for foreground pixels and a zero-displacement target for background pixels.
    \item $\loss_{\text{render}}$ comprises differentiable rendering losses on novel, held-out training views sampled uniformly within the context time window. The loss consists of an $L_1$ term and an LPIPS term~\citep{zhang2018unreasonable}, computed after applying the predicted camera-ISP affine transform. This rendering loss also backpropagates to the sky cubemap decoder. We additionally use an opacity loss to encourage transparency in sky regions so that the sky is represented primarily by the cubemap rather than the 3D Gaussians.
\end{itemize}

In practice, introducing $\loss_{\text{render}}$ at the beginning of training slows convergence. Moreover, rendering at every iteration is too expensive because of the large number of Gaussians. We therefore divide the full training process into a three-stage curriculum:
\begin{itemize}[leftmargin=1.2em]
    \item \textbf{Stage 1: Pretraining.} We reproduced the Depth-Anything-3 pretraining protocol using the depth- and ray-prediction tasks from \cite{lin2025depth} and the data-curation pipeline described in \cite{burzio2026d}. This stage provides a general initialization for the full backbone.
    \item \textbf{Stage 2: Context training.} We enable the sky, depth-and-context, and motion decoders and fully fine-tune the encoder with these heads on the internally curated driving corpus with $\loss_{\text{context}}+\loss_{\text{motion}}$; the context heads receive dense, well-conditioned supervision from LiDAR and auto-labels.
    \item \textbf{Stage 3: GS training.} The encoder is frozen, and the camera-ISP and GS decoders are enabled. Because no ground-truth Gaussian attributes are available, we train these heads only after the geometry has stabilized, allowing rendering gradients to backpropagate through a sensible 3D scaffold. The Gaussians produced by the now-frozen geometry stack are rendered through 3DGUT~\citep{wu20253dgut} and additionally supervised with $\loss_{\text{render}}$.
\end{itemize}
Freezing the encoder in Stage 3 is a deliberate choice: it prevents high-variance rendering gradients from corrupting the well-conditioned geometry features learned in Stage 2 and substantially stabilizes training while the GS attributes are still random.

\subsection{Implementation Details}
\label{sec:impl}
\label{sec:data}

\parahead{Data curation and auto-labeling}
\MethodName{} is trained on $\sim$40K filtered driving clips drawn from NVIDIA's Autonomous Vehicle (AV) data platform.
Each clip contains footage from up to five cameras, with $\sim$300--600 frames per camera at 30\,Hz, along with the corresponding camera calibrations and rig-trajectory data. Each clip also includes LiDAR point-cloud data, from which we export depth-map ground truth.
In addition to the raw sensor data and calibrations, we apply the following steps to obtain auxiliary training data:
\begin{itemize}[leftmargin=1.2em]
    \item \textbf{Semantic segmentation} is obtained by independently applying a state-of-the-art semantic segmentation model~\citep{nurec_website} to every image from each camera.
    \item \textbf{3D bounding boxes} for movable actors are provided by an internal cuboid auto-labeler and tracker using LiDAR data.
    \item \textbf{Dense depth} is obtained by accumulating LiDAR points across the clip and projecting them into each image; missing pixels (sky, ego-car, and occluded surfaces) are masked out of the depth loss. Regions belonging to dynamic actors are accumulated according to their auto-labeled bounding boxes.
\end{itemize}

\parahead{Training infrastructure}
\MethodName{} is implemented in PyTorch, with 3DGUT~\citep{wu20253dgut} serving as the differentiable renderer.
Training uses standard DDP across nodes with an effective per-GPU batch size of 1--2 clips. We train the model with several common camera configurations $V \in \{1,3,5\}$ (front only; front, front-left, and front-right; or front, front-left, front-right, rear-left, and rear-right) and varying temporal frame counts $T \in \{8,12,18\}$.
The learning rate follows a cosine annealing schedule with a linear warmup to a maximum of $10^{-4}$. The complete three-stage training process takes $\sim$6 days on 8 nodes.

\section{Experiments}
\label{sec:experiments}

We evaluate \MethodName{} along three axes: \textit{(i)} novel-view synthesis quality against recent feed-forward baselines on the public Waymo Open Dataset (\Cref{sec:exp_waymo}); \textit{(ii)} reconstruction quality and closed-loop simulation viability on our large-scale internal corpus (\Cref{sec:exp_internal}); and \textit{(iii)} ablations of the main design choices (\Cref{sec:exp_ablation}).

\subsection{Comparison on the Waymo Open Dataset}
\label{sec:exp_waymo}

For reproducibility and direct comparison with recent feed-forward reconstruction methods, we evaluate on the validation split of the Waymo Open Dataset~\cite{sun2020scalability}. Our evaluation follows the protocol used by STORM~\cite{yang2025storm}. Each evaluation sample is a 2-second window containing 20 consecutive frames at 10\,Hz. The model observes four context frames spaced 0.5\,s apart and is evaluated on the held-out target frames. All RGB metrics are computed after resizing the predictions and ground truth to the STORM resolution of $240\times160$ pixels.

\parahead{Baselines}
We compare against representative feed-forward reconstruction methods: DepthSplat~\cite{xu2024depthsplat}, STORM~\cite{yang2025storm}, Depth-Anything-3~\cite{lin2025depth}, and DGGT~\cite{chen2025dggt}. We evaluate the baseline methods using their released checkpoints without further fine-tuning. Because DGGT does not take poses as input, we align its predicted trajectory with the ground-truth trajectory before computing the metrics.

\parahead{Metrics}
For image quality, we report PSNR and SSIM on both the full image and the dynamic region. We define the dynamic region using semantic segmentation masks that select vehicle and movable-object pixels, and we use the same mask for all methods. For geometry, we follow the sparse-LiDAR depth evaluation protocol used in StreetForward~\cite{yu2026streetforward}. We project Waymo LiDAR points into the front camera and evaluate only pixels with valid sparse depth. Because several feed-forward baselines predict depth up to an arbitrary scale, we align each prediction with the LiDAR depth using a per-frame least-squares scale before reporting AbsRel and $\delta_1$ for all valid LiDAR pixels and for the subset in dynamic regions.

\begin{figure}[htbp]
    \centering
    \includegraphics[width=\linewidth]{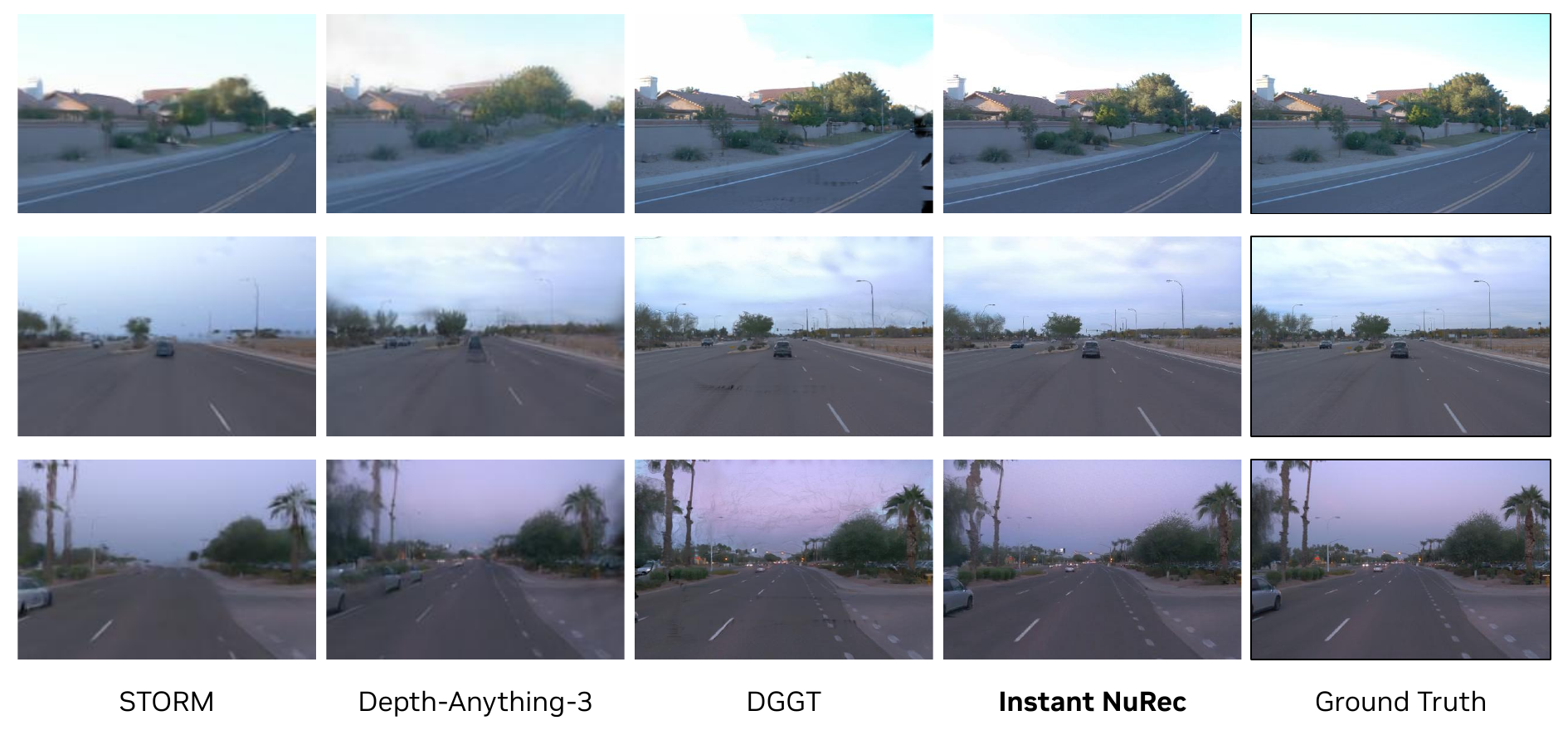}
    \caption{\textbf{Qualitative comparison on the Waymo Open Dataset.} Novel-view synthesis results for held-out target frames from the validation set. Compared with recent feed-forward baselines, \MethodName{} produces sharper imagery and better-preserved thin structures.}
    \label{fig:waymo_qualitative}
\end{figure}

As shown quantitatively in \Cref{tab:waymo} and qualitatively in \Cref{fig:waymo_qualitative}, our model achieves the best RGB reconstruction among the evaluated feed-forward baselines, with even larger margins in dynamic regions. Its depth quality is on par with that of the strongest baselines, if not better on some metrics.
We attribute the improvements mainly to our choice of efficient backbones and more effective training strategies.

\begin{table}[htbp]
\centering
\caption{Image and depth quality comparison on the Waymo Open Dataset. Higher PSNR/SSIM/$\delta_1$ and lower AbsRel are better.}
\label{tab:waymo}
\setlength{\tabcolsep}{10pt}
\footnotesize
\begin{tabular}{lcccccccc}
\toprule
\multirow{2}{*}[-0.5ex]{Method} &
\multicolumn{4}{c}{Full Image} &
\multicolumn{4}{c}{Dynamic Region} \\
\cmidrule(lr){2-5}
\cmidrule(lr){6-9}
& PSNR $\uparrow$ & SSIM $\uparrow$ & AbsRel $\downarrow$ & $\delta_1 \uparrow$
& PSNR $\uparrow$ & SSIM $\uparrow$ & AbsRel $\downarrow$ & $\delta_1 \uparrow$ \\
\midrule
DepthSplat~\cite{xu2024depthsplat} & 22.48 & 0.645 & 0.295 & 0.592 & 18.59 & 0.391 & 0.445 & 0.512 \\
STORM~\cite{yang2025storm} & 21.88 & 0.752 & 0.123 & 0.870 & 19.89 & 0.556 & 0.178 & 0.789 \\
Depth-Anything-3~\cite{lin2025depth} & 20.30 & 0.557 & 0.434 & 0.313 & 17.59 & 0.250 & 0.467 & 0.354 \\
DGGT~\cite{chen2025dggt} & 26.25 & 0.805 & 0.135 & 0.841 & 21.76 & 0.652 & 0.249 & 0.689 \\
\textbf{Ours} & \textbf{28.26} & \textbf{0.859} & \textbf{0.076} & \textbf{0.937} & \textbf{24.93} & \textbf{0.793} & \textbf{0.085} & \textbf{0.922} \\
\bottomrule
\end{tabular}
\end{table}

\subsection{Comparisons and Applications on Internal Data}
\label{sec:exp_internal}

Public driving datasets are relatively small and cover only a limited range of conditions.
To demonstrate real-world applicability, we evaluate on a held-out validation split of our internal corpus and assess reconstruction quality and simulator integration in settings that public benchmarks rarely capture, including non-pinhole cameras, longer sequences (up to 30\,s), and complex driving scenarios.
Throughout this section, our baseline is NVIDIA NuRec~\citep{nurec_website,carla_nurec_docs}, a per-scene-optimized 3DGS reconstruction that requires roughly $75$\,min of optimization per scene and serves as our performance reference for production-quality reconstruction.

\begin{figure}[htbp]
    \centering
    \includegraphics[width=\linewidth]{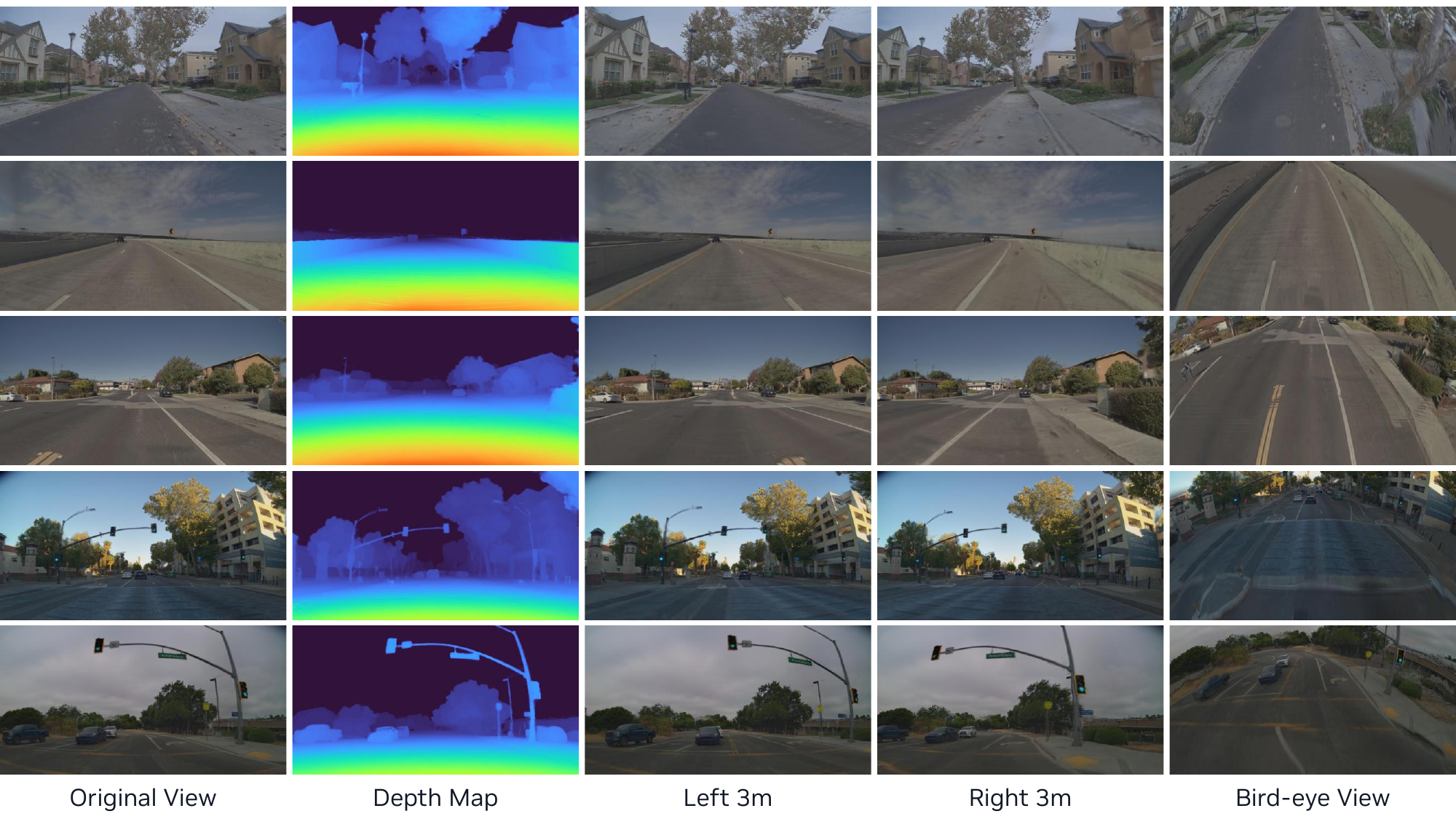}
    \caption{\textbf{Single-camera reconstruction.} From a single front-camera input, \MethodName{} reconstructs a navigable 3DGS scene in one forward pass. The re-posed and top-down renderings remain geometrically consistent and preserve thin structures such as poles and traffic lights.}
    \label{fig:ncore_1cam}
\end{figure}

\begin{figure}[htbp]
    \centering
    \includegraphics[width=\linewidth]{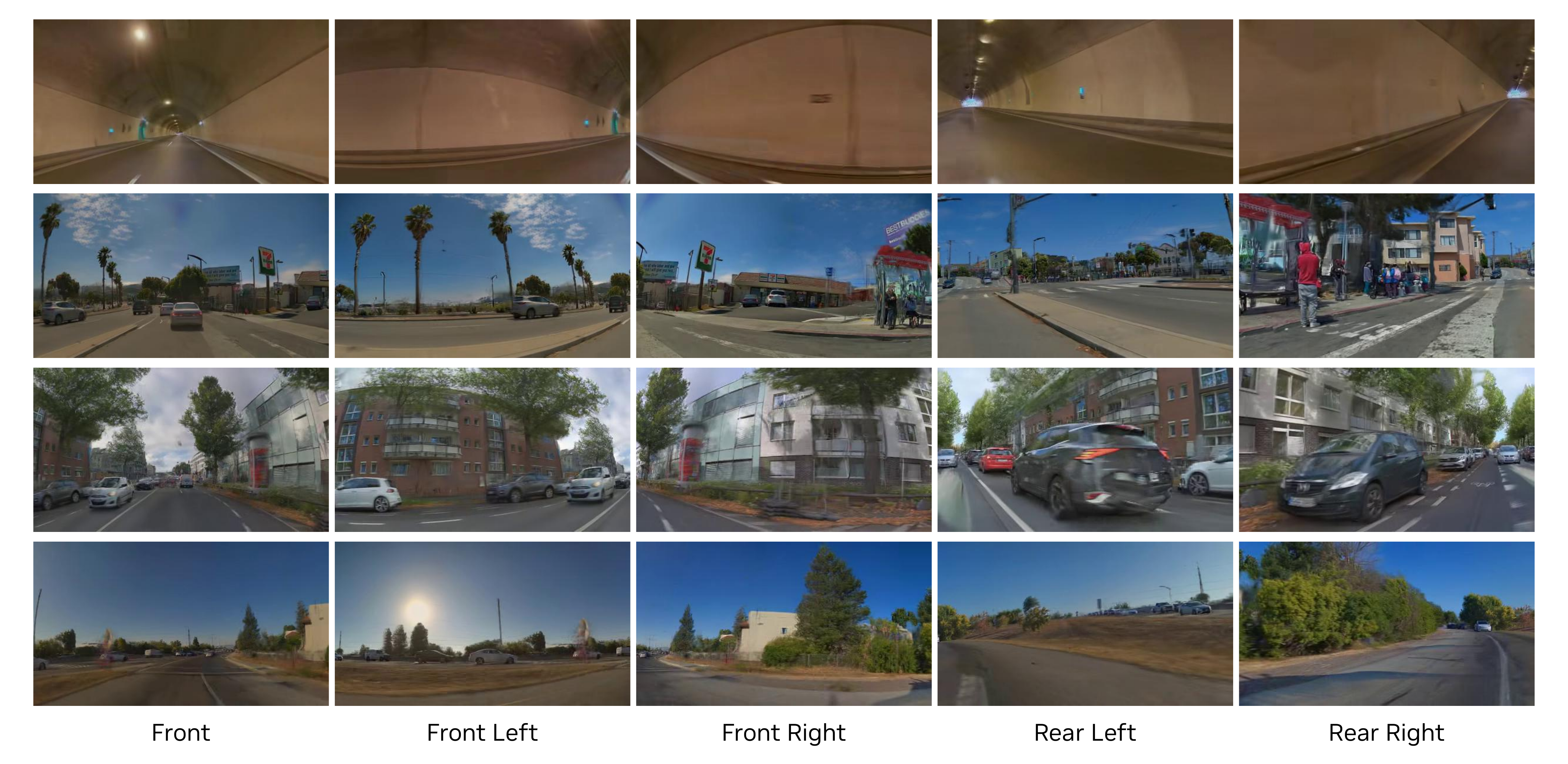}
    \caption{\textbf{Multi-camera surround-view reconstruction.} From a five-camera input rig, \MethodName{} jointly reconstructs the full surround view. Each row shows a different scene, and the rendered views remain consistent in appearance and geometry across overlapping cameras.}
    \label{fig:ncore_5cam}
\end{figure}

\parahead{Reconstruction quality}
\Cref{fig:ncore_1cam} shows single-camera reconstructions in which re-posed and bird's-eye views remain geometrically consistent and preserve thin structures across diverse scenes.
\MethodName{} also reconstructs the full multi-camera surround view in a single forward pass: \Cref{fig:ncore_5cam} shows the result across all five cameras (front, front-left, front-right, rear-left, rear-right), with consistent appearance and geometry across the rig in scenes ranging from tunnels to dense urban streets.

\begin{figure}[htbp]
    \centering
    \includegraphics[width=\linewidth]{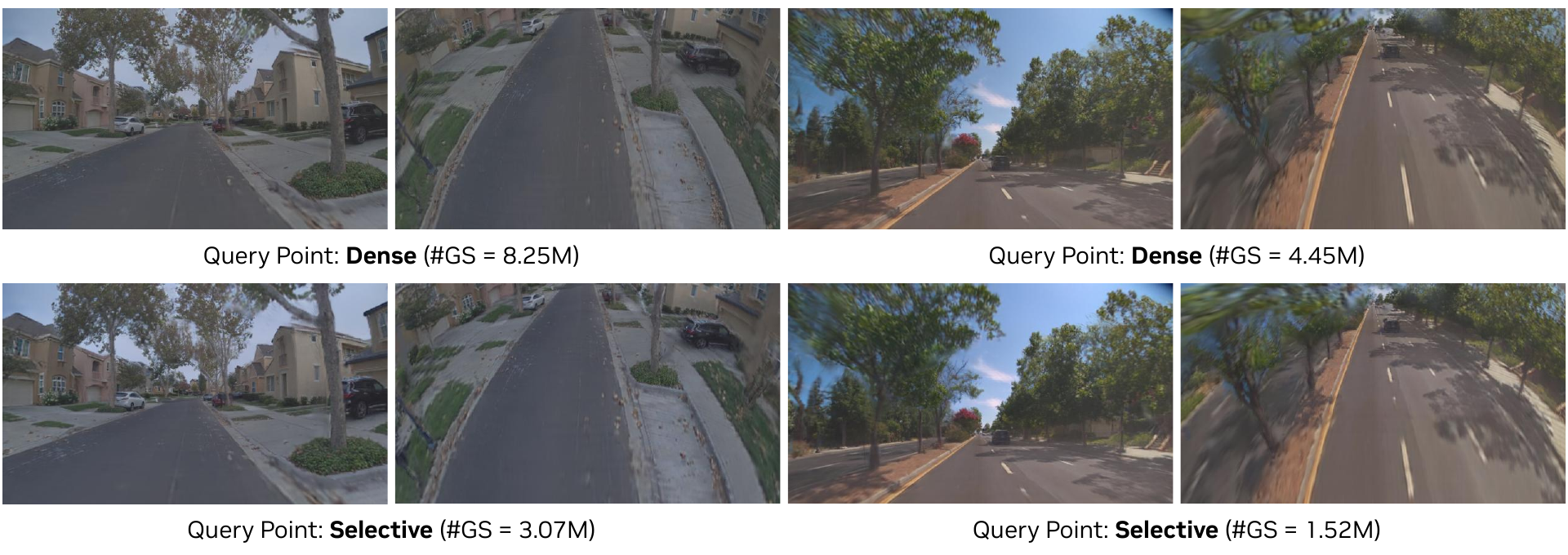}
    \caption{\textbf{Query-point selection.} Comparison of the \textbf{Dense} and \textbf{Selective} query-point strategies. The Selective strategy produces far fewer Gaussians (\#GS) while maintaining comparable reconstruction quality.}
    \label{fig:ncore_pq}
\end{figure}

\begin{figure}[htbp]
    \centering
    \includegraphics[width=\linewidth]{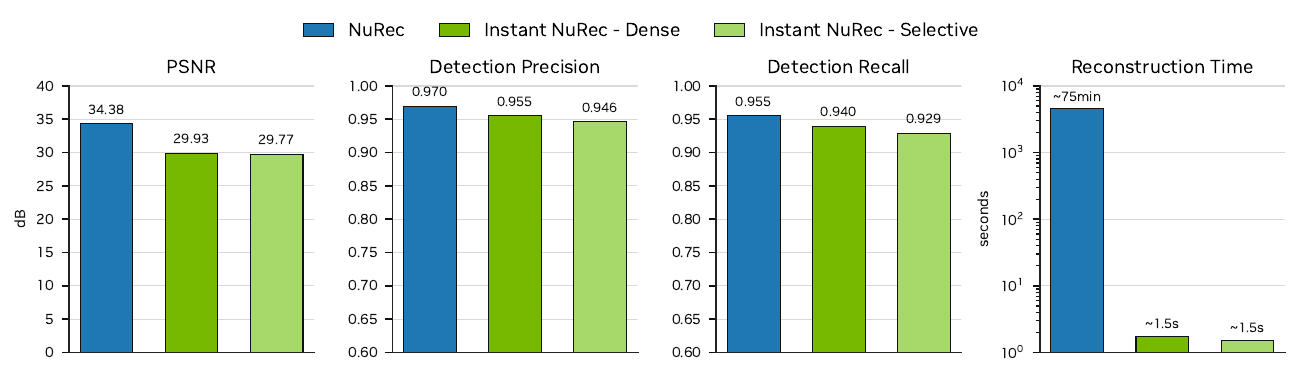}
    \caption{\textbf{Quantitative reconstruction comparison on internal data.} \MethodName{} (Dense and Selective) versus the per-scene-optimized NuRec baseline on PSNR, downstream 3D detection precision and recall, and reconstruction time. \MethodName{} reaches comparable appearance and detection quality while reconstructing a scene in $\sim$1.5\,s instead of $\sim$75\,min.}
    \label{fig:ncore_metrics}
\end{figure}

As described in \Cref{sec:method}, \MethodName{} provides two query-point strategies, \textbf{Dense} and \textbf{Selective}, that trade Gaussian budget for reconstruction quality.
On our validation set, the Selective strategy cuts the Gaussian budget by roughly $3\times$---from $\sim$351K per context view for Dense to $\sim$120K ($\sim$6.3M to $\sim$2.2M per 18-frame scene)---with only marginal losses in image and depth accuracy, as shown in \Cref{fig:ncore_pq}.
\Cref{fig:ncore_metrics} further compares both variants with NuRec: both reconstruct a scene in $\sim$1.5\,s versus $\sim$75\,min for NuRec, a speedup of more than three orders of magnitude, while trailing only modestly in appearance quality.

We further evaluate reconstruction quality by applying an in-house 3D detection pipeline to the reconstructed scenes. We render each scene from novel viewpoints and compare the resulting detection precision and recall with those obtained from ground-truth images captured by the ego cameras. This evaluation checks that reconstruction inaccuracies do not bias downstream policies.
Dense remains the stronger reconstruction variant, while Selective incurs a modest reduction in appearance and detection metrics while using roughly one-third of the Gaussian budget. Selective is therefore useful when memory or rendering cost is the primary constraint, while Dense provides the highest-quality feed-forward reconstruction.

\begin{figure}[htbp]
    \centering
    \includegraphics[width=\linewidth]{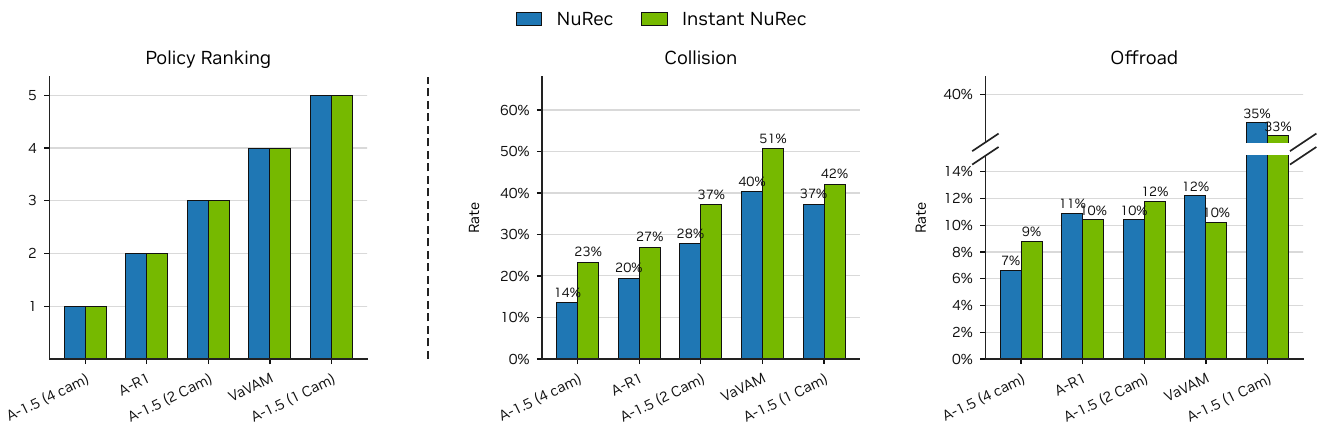}
    \caption{\textbf{Closed-loop comparison between NuRec and \MethodName{} across five policy configurations.}
    Each bar pair compares the same policy under NuRec (blue) versus \MethodName{} (green), averaged over 140 scenes $\times$ 6 trials.
    Despite requiring only a single forward pass, \MethodName{} reproduces the same policy ranking as NuRec, validating it as a reliable simulator for policy comparison.}
    \label{fig:closed_loop}
\end{figure}

\parahead{Closed-loop simulation viability}
A key question for any feed-forward reconstruction method is whether the reconstruction produced by a single forward pass is sufficient for reliable closed-loop policy simulation.
To test this, we compare \MethodName{} with NuRec by running the same set of policies through the same AlpaSim~\citep{alpasim_2025} closed-loop stack and changing only the reconstructions.
Our evaluation set contains 140 scenes. Each scene is rolled out for 20\,s in closed loop, with policy replanning every 500\,ms and six independent trials per scene.
We simulate five driving policy configurations: VaVAM~\citep{vavam2025}, Alpamayo~R1~\citep{nvidia2025alpamayo} (A-R1), and the 1-camera, 2-camera, and 4-camera variants of Alpamayo~1.5~\citep{nvidia2025alpamayo} (A-1.5).
We evaluate the policies using the average of two key metrics: \emph{Collision (Front + Lateral + Rear)} and \emph{Offroad}.
As shown in \Cref{fig:closed_loop}, the policy ranking under \MethodName{} is identical to that under NuRec, meaning that policy selection decisions made with \MethodName{} would match those made with the more expensive per-scene reconstruction.
This makes the method a reliable substitute for per-scene optimization in closed-loop policy evaluation, at a fraction of the reconstruction cost.

\subsection{Ablation Study}
\label{sec:exp_ablation}

\begin{figure}[htbp]
    \centering
    \includegraphics[width=\linewidth]{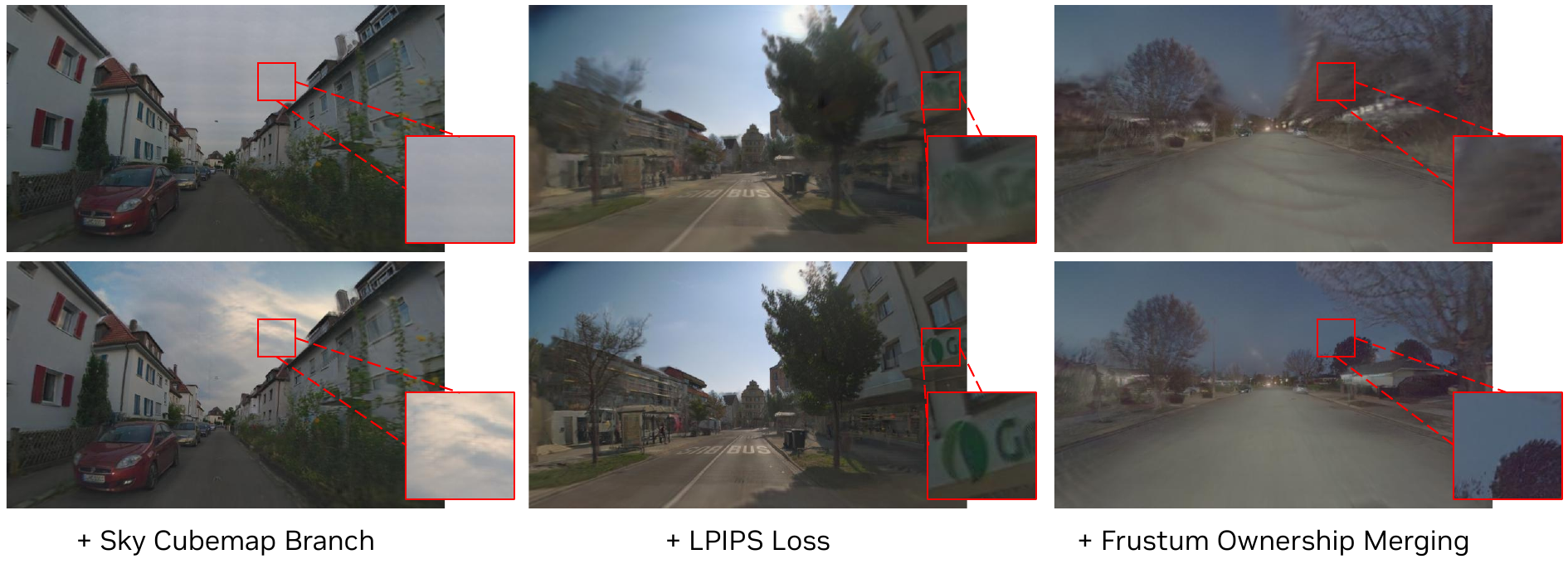}
    \caption{\textbf{Qualitative ablation.} Effects of the sky cubemap branch, LPIPS loss, and frustum-ownership merging strategy. In each comparison, the top row omits the named component, while the bottom row shows the full model.}
    \label{fig:ablation}
\end{figure}

We ablate the main design choices of \MethodName{} on the validation set introduced in \Cref{sec:exp_internal}. We summarize the results quantitatively in \Cref{tab:ablation} and qualitatively in \Cref{fig:ablation}; each variant removes one component to show its individual effect on reconstruction quality.

\parahead{Training strategy}
We implement a single-stage variant that trains Stage~2 and Stage~3 jointly rather than following our sequential schedule.
This variant takes longer to train and requires more iterations to converge, possibly because of the additional complexity introduced by gradients from the renderer. Nevertheless, its final metrics remain worse (\Cref{tab:ablation}), so we retain the multi-stage schedule.

\parahead{Sky model}
We replace the sky cubemap branch with an MLP sky model similar to STORM~\cite{yang2025storm}.
Because of the network's low-frequency inductive bias, this variant fails to recover high-frequency sky textures such as clouds and the sun (\Cref{fig:ablation}).

\parahead{Loss terms}
We ablate the two auxiliary losses in turn. Removing the LPIPS loss makes the reconstructions noticeably blurrier, while removing the depth loss causes the model to overfit to appearance and degrade geometry, as reflected in the quantitative depth metrics.

\parahead{Primitive merging strategy}
We replace frustum-ownership merging with a naive concatenation of Gaussians without pruning.
Without this pruning, far-field Gaussians from the outer region of one chunk can appear as floaters inside neighboring chunks, where they compete with better-conditioned local primitives and degrade quality, especially for longer sequences.

\begin{table}[htbp]
    \centering
    \caption{\textbf{Ablations on internal data.} Reconstruction quality when each component is removed.}
    \label{tab:ablation}
    \setlength{\tabcolsep}{24pt}
    \footnotesize
    \begin{tabular}{lcccc}
        \toprule
        Variant & PSNR $\uparrow$ & SSIM $\uparrow$ & AbsRel $\downarrow$ & $\delta_1 \uparrow$ \\
        \midrule
        \MethodName{} (full)                & \textbf{29.93} & \textbf{0.793} & 0.069          & 0.950          \\
        \midrule
        \quad single-stage training         & 27.65          & 0.751          & 0.077          & \textbf{0.955} \\
        \quad sky MLP instead of cubemap & 26.73          & 0.692          & \textbf{0.068} & 0.950          \\
        \quad w/o LPIPS loss                & 26.81          & 0.705          & 0.073          & 0.934          \\
        \quad w/o depth loss                & 28.92          & 0.782          & 0.103          & 0.862          \\
        \quad w/o frustum-ownership merging & 26.10          & 0.648          & 0.098          & 0.891          \\
        \bottomrule
    \end{tabular}
\end{table}

\section{Discussion and Conclusion}
\label{sec:conclusion}

\parahead{Limitations}
\MethodName{} faces an inherent tension between the number of Gaussians and reconstruction quality: matching the quality of per-scene optimization can require many primitives, while a smaller budget under-samples thin structures. Test-time training that refines the predicted Gaussians on the input clip is a promising way to close this gap.
Generalization is also limited by the training corpus, so rigs that differ substantially from those seen in training (e.g., low-mounted or fisheye-only setups) currently require fine-tuning.
Finally, the three-keyframe piecewise-linear actor trajectories cannot capture sub-second non-rigid motion, such as pedestrian articulation. Denser keyframes could address this limitation.
Separately, a streaming formulation in which past reconstructions condition the current forward pass is a promising direction for scaling to long logs.
\MethodName{} also covers only the camera modality; the companion work Instant LiDAR~\citep{kang2026instantlidar} applies the same feed-forward recipe to LiDAR, reconstructing multi-scan sweeps into reusable Gaussian assets for LiDAR simulation.

\parahead{Conclusion}
\MethodName{} demonstrates that the cost of high-quality driving-scene reconstruction can be amortized into a single feed-forward pass. It turns a short multi-view log into a layered, fully simulatable 3DGS world consisting of static and dynamic Gaussians, a sky cubemap, and per-camera ISP corrections, directly consumable by NuRec and AlpaSim.
By approaching per-scene quality at a cost that is orders of magnitude lower, we hope that \MethodName{} will lower the barrier to using neural reconstruction at the fleet scale that AD development now demands.

\clearpage
{
\small

\bibliographystyle{unsrtnat}
\bibliography{neurips_2025}

}

\clearpage
\appendix
\section{Contributors}
\label{sec:contributors}

\parahead{Research}
Jiahui Huang, Jiawei Ren, Michal Tyszkiewicz, Xin Kang, Seung Wook Kim, Shengyu Huang, Laura Leal-Taixe$^\ddagger$, Zan Gojcic$^\ddagger$, Sanja Fidler$^\ddagger$.

\parahead{Engineering}
Bjoern Haefner, Michael Shelley, Ning Xu, Qi Wu, Janick Martinez Esturo, Nick Schneider$^\ddagger$.

\section{Acknowledgments}

We greatly appreciate the contributions of the following individuals:

Alessandro Burzio, Alex Perec, Bingxin Ke, Daniel Dworakowski, Despoina Paschalidou, Emmanuel Attia, Jun Gao, Katarina Tothova, Lei Zhang, Lucrezia Shen, Murat Arar, Naveen Kumar Rai, Nicolas Moenne-Loccoz, Rodolfo Lima, Sangeetha Grama Srinivasan, Sean Pieper, Sergio Agostinho, Sherwin Bahmani, Shikhar Solanki, Sipeng Zhang, Tianchang Shen, Tobias Fischer, Weihua Zhang, Xuanchi Ren, Yixin Cao.

\begingroup
\renewcommand{\thefootnote}{\ensuremath{\ddagger}}
\footnotetext{Leadership.}
\endgroup

\end{document}